# Universal spectral shapes of nonlinear light scattering from laser-driven dilute-gas density gratings

Aleksei Zheltikov

Institute for Quantum Science and Engineering, Department of Physics and Astronomy, Texas A&M University, College Station, Texas 77843; zheltikov@tamu.edu

Operating within the framework of the collisionless Boltzmann equation, we identify universal spectral profiles inherent in stimulated and coherent light scattering from laser-driven dilute-gas density gratings and derive physically intuitive solutions for such spectral shapes, expressing them in a manifestly causal closed form that satisfies the Kramers – Kronig relations and fulfills the fluctuation – dissipation theorem. Benchmarking against gold-standard experiments on dilute-gas stimulated light scattering verifies a high descriptive and explanatory power of these solutions and demonstrates their ability to isolate universality motifs in nonlinear scattering spectra as the key to fully understanding the physics behind dilute-gas nonlinear light scattering. As one of important outcomes, such analysis reveals the physical content of the bandwidth of dilute-gas nonlinear scattering spectra as an expression and a quantifier of the phase-mixing time, i.e., the time required for free-streaming gas particles with a mean thermal velocity to traverse the distance equal to the spacing of the laser-driven density grating.

## 1. Introduction

Among multitudes of physically significant spectral lineshapes known in spectroscopy, statistical physics, and signal analysis, the Gaussian spectrum stands out as an ultimate benchmark, whose overarching agency is guaranteed by the central limit theorem [1 – 5]. In stochastic dynamics, Gaussian lineshapes emerge as a universal, model-nonspecific collisionless limit for the spectrum of uncorrelated fluctuations, such as density fluctuations in a gas of Maxwellian-distributed particles. As some of its most significant properties, the Gaussian spectral density and its Fourier-transform time-domain Gaussian counterpart minimize the time – bandwidth product in spectral analysis, achieve the lower bound of the Gabor limit in information processing, and provide minimum uncertainty in quantum mechanics [6 – 8]. In stochastic dynamics, the Gaussian spectrum expresses a universal, model-nonspecific collisionless limit for the spectrum of uncorrelated fluctuations [1, 2, 6, 7], such as density fluctuations in a gas of Maxwellian-distributed free-streaming, ballistic particles. In atomic and molecular spectroscopy, this limit is manifested as purely inhomogeneous spectral line broadening, readily understood in terms of the Doppler effect [3, 5, 9].

In this study, we examine how the Gaussian spectra inherent in spontaneous density fluctuations in dilute gases transform into much less intuitive, yet physically significant and insightful spectra of stimulated and coherent scattering. Although our treatment is applicable to a broad range of nonlinear processes, our specific focus here is on coherent and stimulated light scattering from laser-driven density fluctuations, including coherent and stimulated Brillouin and Rayleigh scattering [10 – 13], as archetypal processes of nonlinear scattering that play a prominent role in spectroscopy and laser technologies. Some of the most significant applications of these processes include nonlinear spectroscopy [11, 14 – 18], fiber-optic sensing [19 – 21], bioimaging [22 – 29], tunable all-optical delay and slow-light generation [30 – 32], and diagnostics of high-speed flows and fluid dynamics [33 – 48]. Alongside its Raman counterpart, stimulated Brillouin scattering, one of the processes examined in this study, is among the most important parametric instability processes in laser-produced plasmas [49 – 53], including plasmas of laser-driven inertial confinement fusion [54 – 57]. In a suitably engineered laser–plasma interaction setting, on the other hand, SBS can provide a mechanism for an efficient pulse amplification and compression [58 – 66], opening routes toward sources of high-intensity subfemtosecond x-ray pulses [67]. Prominent applications of SBS in advanced laser technologies include creation of phase-conjugation mirrors [68 – 71], narrow-linewidth lasers [72 – 77], as well as compressors, beam combiners, and amplifiers of high-energy laser pulses [78 – 84], including gas-phase pulse compressors and beam combiners for the latest generation of high-energy laser facilities for inertial confinement fusion [85 – 89].

In solids, liquids, and high-pressure gases, low-frequency light scattering is adequately understood [10 – 14, 90] in terms of scattering by dielectric permittivity variations – random in the spontaneous scattering or regular, via laser-induced density gratings, in coherent and stimulated regimes. A standard framework for the analysis of such a process in solids and fluids is based on a set of hydrodynamic equations consisting of the Navier – Stokes equation with an electrostrictive force, continuity equation, and the heat-conduction equation with a source term including absorption of laser radiation [11 – 13]. In moderate-pressure gases, however, where the mean free path of gas particles is longer than the effective wavelength of the density wave transferring the energy of the laser driver to the scattered field, the entire picture of light – sound scattering breaks, as indicated by drastic changes in the properties of scattered light and the shape of the gain spectrum [33 – 36, 91, 92]. In this regime, light scattering can no longer be viewed as scattering by continuous fluid but should be treated instead as scattering by an ensemble of independent, free-streaming particles. This regime of scattering, often referred to as the kinetic or dilute-gas regime, has long been a subject of much interest in the context of applications of stimulated and coherent scattering in nonlinear spectroscopy of atomic and molecular gases [33 – 36, 43 – 46, 91, 92], laser-based gas-flow diagnostics [37 – 40, 47, 48], and, most recently, pulse compression in laser drivers for inertial fusion [88, 89, 93].

As the gas ceases to be a continuous fluid in a low- or moderate-pressure regime, it can no longer be described hydrodynamically but should be treated using suitably adapted kinetic equations. Specifically, the properties of spontaneous low-frequency light scattering are adequately understood [94 – 100] in terms of the spectrum of density fluctuations found by solving the Boltzmann equation with a suitable linearization of the collision operator, e.g., using the single-relaxation-time linear Bhatnagar – Gross – Krook (BGK) approximation [101] or models with multiple independent relaxation rates, such as the Shakhov model [102, 103] or the Tenti models [104 – 106]. Extension of this framework to stimulated and coherent regimes of light – matter interaction has been achieved by adding field-dependent source terms to the linearized Boltzmann equation (LBE) in one of these approximations. Such generalizations have been shown to provide an accurate description of experiments on coherent Rayleigh [45] and coherent Rayleigh – Brillouin scattering [38 – 40], as well as experiments on stimulated low-frequency light scattering [93]. However, because these solutions rederive the spectrum of density fluctuations from the field-modified LBE, expressing this spectrum as the ratio of complex polynomials of the plasma dispersion function, they do not always provide a clear, physically intuitive insight into how light-driven density fluctuations give rise to light-intensity-dependent gain and coherence buildup in stimulated and coherent regimes of light scattering. Among the fundamental physical questions that remain to be addressed are the questions of how the causality of density fluctuations is manifested in the spectra of stimulated scattering and how the universality of Doppler-broadened spectra emerging in spontaneous scattering off density fluctuations translates into the spectra of stimulated and coherent scattering. Standing out among unanswered practical questions is the question of whether the kinetic regime of stimulated scattering from laser-driven density gratings can provide any practically significant gain enhancement for efficient pulse compression in high-power laser beamlines, including laser systems for inertial fusion.

Here, operating with the collisionless Boltzmann equation, we derive closed-form solutions for the universal spectral profiles inherent in stimulated and coherent light scattering from laser-driven density gratings in dilute gases. We show that these profiles can be expressed in terms of the first derivative of the Faddeeva function. When cast in this form, the spectra of nonlinear scattering are manifestly causal, satisfying the Kramers – Kronig relations and fulfilling the fluctuation – dissipation theorem. As one of their key properties, the bandwidth of these spectra is defined by the characteristic phase-mixing time, i.e., the time required for gas particles with a mean thermal velocity to traverse the distance equal to the spacing of the laser-driven density grating.

## 2. Electrostriction-induced nonlinear polarization and stimulated scattering

We consider a standard setting of stimulated and coherent light scattering [11 – 15] in which a laser field,

$$E(\mathbf{r},\mathrm{t}) = E_p \exp\left[i\left(\mathbf{k}_p \cdot \mathbf{r} - \omega_p t\right)\right] + E_s \exp[i(\mathbf{k}_s \cdot \mathbf{r} - \omega_s t)] + c.c., \qquad (1)$$

consisting of counterpropagating pump and Stokes/seed laser fields with field amplitudes $E_p$ and $E_s$, wave vectors $\mathbf{k}_p$ and $\mathbf{k}_s$, and central frequencies $\omega_p$ and $\omega_s$, induces a ponderomotive potential

$$U_{\mathrm{p}}(\mathbf{r},\mathrm{t}) = -\frac{\alpha}{2}\overline{[E(\mathbf{r},\mathrm{t})]^2}, \qquad (2)$$

where $\alpha$ is the polarizability of gas particles.

This potential gives rise to a ponderomotive force

$$\mathbf{F}(\mathbf{r},\mathrm{t}) = -\nabla U_{\mathrm{p}}(\mathbf{r},\mathrm{t}) = \frac{\alpha}{2}\nabla\overline{[E(\mathbf{r},\mathrm{t})]^2}. \qquad (3)$$

which pushes gas particles toward regions of higher field strength, thus inducing a change $\delta\rho(\mathbf{r},t)$ in gas density $\rho(\mathbf{r},t)$. The resulting change $\delta\varepsilon(\mathbf{r},t)$ in dielectric permittivity $\varepsilon(\mathbf{r},t)$ couples the pump and Stokes fields via the nonlinear polarization, which, in the first order in $\delta\rho(\mathbf{r},t)$, i.e., with

$$\delta\varepsilon(\mathbf{r},t) \approx \left(\frac{\partial\varepsilon}{\partial\rho}\right)\delta\rho(\mathbf{r},t), \qquad (4)$$

is given by [11, 12]

$$P^{(3)}(\mathbf{r},\mathrm{t}) = \frac{\delta\varepsilon(\mathbf{r},t)}{4\pi}E(\mathbf{r},t) = \frac{1}{4\pi}\left(\frac{\partial\varepsilon}{\partial\rho}\right)\delta\rho(\mathbf{r},t)E(\mathbf{r},t). \qquad (5)$$

When treated in the first order in $\delta\rho$, this nonlinear polarization is cubic in the laser field $E(\mathbf{r},t)$ because $\delta\rho$ is quadratic in $E(\mathbf{r},t)$. Nonlinear coupling between the pump and Stokes fields and nonlinear-polarization wave opens a channel for energy transfer from the pump to the Stokes field, as described by the standard coupled-wave equations for the slowly varying amplitudes $E_p(\mathbf{r}, \mathrm{t})$, $E_s(\mathbf{r}, \mathrm{t})$, and $\delta\rho(\mathbf{r}, t)$ [11, 12, 90]. As a part of this process, the pump and Stokes field are coupled via the third-order Fourier-space nonlinear polarization at frequency $\omega_s$,

$$P^{(3)}(\omega_s) = \frac{1}{4\pi}\left(\frac{\partial\varepsilon}{\partial\rho}\right)\delta\rho_\Omega^* E_p = \frac{\gamma_e}{4\pi\rho}\delta\rho_\Omega^* E_p = \chi^{(3)}(\Omega)\left|E_p\right|^2 E_s, \tag{6}$$

where $\gamma_e = \rho\left(\frac{\partial\varepsilon}{\partial\rho}\right) = \varepsilon - 1 = 4\pi\mathcal{N}\alpha$ is the electrostrictive coefficient, $\mathcal{N}$ is the number density of gas particles, $\chi^{(3)}(\Omega)$ is the third-order nonlinear susceptibility [10 – 12, 14, 14], and $\delta\rho_\Omega = \delta\rho(\mathbf{K}, \Omega)$ is the Fourier amplitude of the laser-driven grating $\delta\rho(\mathbf{r}, t)$ with frequency $\Omega = \omega_p - \omega_s$ and wave vector $\mathbf{K} = \mathbf{k}_p - \mathbf{k}_s$. Because, by the very nature of this process, $\Omega \ll \omega_p, \omega_s$ and $k_p = \left|\mathbf{k}_p\right| \approx k_s = |\mathbf{k}_s|$, $K = |\mathbf{K}| \approx 2k_p \sin(\theta/2)$, where $\theta$ is the angle between $\mathbf{k}_p$ and $\mathbf{k}_s$. Specifically, for counterpropagating pump and Stokes beams, $\theta = \pi$, we find $K \approx 2k_p$.

With the pump and Stokes fields phase-matched as needed for the maximum efficiency of stimulated scattering, the coupled equations for $E_p(\mathbf{r}, \mathrm{t})$, $E_s(\mathbf{r}, \mathrm{t})$, and $\delta\rho(\mathbf{r}, t)$, solved in the limit of undepleted pump, lead to the signature exponentially growing solution for the Stokes field intensity [10 – 12, 90],

$$I_s(z) = I_s(z = 0)\exp\left[g_s(\Omega) I_\mathrm{p} z\right], \tag{7}$$

with constant pump intensity $I_\mathrm{p}$, coordinate $z$ chosen along $\mathbf{k}_s$, and gain factor

$$g_s(\Omega) = -\frac{8\pi^2\omega_s}{n^2c^2}\mathrm{Im}\left[\chi^{(3)}(\Omega)\right]. \tag{8}$$

Finding the gain factor $g_s(\Omega)$ requires a specific physical model connecting the field-induced density change $\delta\rho(\mathbf{r}, t)$ to thermodynamic parameters of the medium. In standard settings of stimulated Brillouin scattering [11, 12, 14, 15, 90], such a model is formulated in terms of the Navier – Stokes equation in conjunction with the continuity equation for $\rho(\mathbf{r}, t)$, providing equations of motion for the gas-density and temperature variations. In this canonical picture, SBS is adequately understood as a light–sound interaction in which an intense optical pump couples to a Stokes field via an intermediary of a yoked acoustic wave, leading to Stokes-field amplification.

In moderate- or low-pressure gases, however, stimulated Stokes amplification is also observed in settings in which the spacing $l_g = 2\pi/K$ of the density grating needed to satisfy momentum conservation $\mathbf{k}_p = \mathbf{k}_s + \mathbf{K}$ is smaller, $l_g < l_0$, than the mean free path $l_0$ of gas particles. Although generation of sound waves with such $K$ is not possible, the Stokes field, as kinetic analysis shows [35, 36, 91, 92], can still be amplified. Such amplification is understood in terms of stimulated scattering off the light-induced density grating. Because $l_0 = v_0/\nu_c$, where $v_0$ is the mean velocity of gas particles and $\nu_c$ is the collision frequency, the ratio $l_g/l_0$ can be expressed as $l_g/l_0 = 2\pi y$, where $y = \nu_c/(Kv_0)$. As a useful reference, for a backward stimulated scattering ($\theta = \pi$, $K \approx 2k_p$) of $\lambda_p \approx 266$ nm laser radiation in 1 atm of argon at room temperature, the $y$ parameter is estimated as $y \approx 0.26$, indicating the kinetic regime of light scattering.

## 3. Kinetic treatment

To understand the properties of density fluctuations behind spontaneous, stimulated, and coherent light scattering, we follow the established kinetic approach [38, 40, 45, 107], operating with the Boltzmann equation

$$\left(\frac{\partial}{\partial t} + \mathbf{v}\cdot\boldsymbol{\nabla}\right) f + \frac{\mathbf{F}}{m}\cdot\boldsymbol{\nabla}_{\mathbf{v}} f = \mathcal{C}[f] \tag{9}$$

for the distribution density function $f \equiv f(\mathbf{r}, \mathbf{v}, t)$ in an ensemble of particles with mass $m$ close to its equilibrium in the presence of the laser-driven ponderomotive force $\mathbf{F} = \mathbf{F}(\mathbf{r}, \mathrm{t})$ [Eq. (3)] with a collision operator $\mathcal{C}[f]$.

The solution to Eq. (9) is sought for perturbatively, with $f(\mathbf{r}, \mathbf{v}, t)$ expanded as

$$f(\mathbf{r}, \mathbf{v}, t) = \rho_0 f_0(\mathbf{r}, \mathbf{v}, t) + f_1(\mathbf{r}, \mathbf{v}, t) \tag{10}$$

about the equilibrium stationary distribution $f_0(\mathbf{r}, \mathbf{v}, t)$ with nonperturbed density $\rho_0$.

Substituting Eq. (10) into Eq. (9) yields an equation for the small correction $f_1 \equiv f_1(\mathbf{r}, \mathbf{v}, t)$:

$$\left(\frac{\partial}{\partial t} + \mathbf{v}\cdot\boldsymbol{\nabla}\right) f_1 + \frac{\rho_0}{m}\mathbf{F}\cdot\boldsymbol{\nabla}_{\mathbf{v}} f_0 = \mathcal{C}_1[f_1], \tag{11}$$

with a suitable linearization of the collision operator $\mathcal{C}_1[f_1] = \mathcal{C}[\rho_0 f_0 + f_1] - \mathcal{C}[\rho_0 f_0]$.

The Boltzmann equation with a suitably linearized collision operator without an external force term [Eqs. (9), (11) with $\mathbf{F} = 0$] provides an established framework for the analysis of spontaneous density fluctuations [94 – 106]. As one of their most prominent features, large-$y$ density fluctuation spectra display well-resolved Lorenztian-shape Brillouin sidebands [94 – 99], arising from acoustic modes of density fluctuations, displaced symmetrically about the central peak, corresponding to Rayleigh scattering, caused by thermal fluctuations. This Rayleigh – Brillouin triplet is well-known in spectroscopy of light scattering and is broadly used for the diagnostics of gas flows and analysis of fluid dynamics [37 – 40, 42 – 48].

As $y$ decreases, the Brillouin sidebands tend to merge with the central, Rayleigh peak, forming in the $y < 1$ regime a broad structureless spectral feature centered at $\Omega = 0$. In the $y \to 0$ limit, $S(\Omega)$ becomes Gaussian,

$$\lim_{y \to 0} S(\Omega) = \pi^{-1/2}\Omega_0^{-1} \exp(-\,\Omega^2/\Omega_0^2), \tag{12}$$

with frequency-independent constant $\Omega_0$, as a general property of spontaneous density fluctuations, identified already in early kinetic-theory studies of density correlations in fluids [94].

## 4. Spontaneous density fluctuations in a collisionless gas

To gain better insights into the properties of spontaneous and laser-driven density fluctuations in the $y \ll 1$ regime, we resort to the collisionless Boltzmann equation,

$$\left(\frac{\partial}{\partial t} + \mathbf{v} \cdot \boldsymbol{\nabla}\right) f + \frac{\mathbf{F}}{m} \cdot \boldsymbol{\nabla}_{\mathbf{v}} f = 0. \tag{13}$$

In plasma physics, this equation is often solved jointly with the Maxwell equations [108 – 119]. The resulting set of equations is broadly referred to, in the plasma physics context, as the Vlasov equations.

Spontaneous density fluctuations in a collisionless gas are described by Eq. (13) without an external force,

$$\left(\frac{\partial}{\partial t} + \mathbf{v} \cdot \boldsymbol{\nabla}\right) f = 0. \tag{14}$$

The solution to this equation is $f(\mathbf{r}, \mathbf{v}, t) = f(\mathbf{r} - \mathbf{v}t, \mathbf{v}, 0)$. Position-space correlation functions are found by solving Eq. (14) for a particle from Maxwellian distribution localized at $\mathbf{r} = 0$ at time $t = 0$, i.e., with the initial condition

$$f(\mathbf{r}, \mathbf{v}, t = 0) = f_0(\mathbf{v})\delta(\mathbf{r}). \tag{15}$$

Eq. (14) subject to the initial condition (15) leads to the following solution for the position-space density correlation function for $t > 0$:

$$\langle \delta\rho(\mathbf{r}, 0)\delta\rho(\mathbf{r}', t)\rangle = m\rho_0 \left(\frac{m}{2\pi k_B T t^2}\right)^{3/2} \exp\left(-\frac{m|\mathbf{r}'-\mathbf{r}|^2}{2k_B T t^2}\right), \tag{16}$$

such that $\langle \delta\rho(\mathbf{r}, 0)\delta\rho(\mathbf{r}', 0)\rangle = m\rho_0\delta(\mathbf{r}' - \mathbf{r})$

Transforming to the wave-vector space via $\delta\rho_{\mathbf{k}}(t) = \int \delta\rho(\mathbf{r}, t) \exp(-i\mathbf{k} \cdot \mathbf{r})\, d^3r$, we find

$$R_k(t) = \frac{1}{V}\langle \delta\rho_{\mathbf{k}}(t)\delta\rho_{-\mathbf{k}}(0)\rangle = m\rho_0 \int f_0(\mathbf{v}) \exp(-i\mathbf{k} \cdot \mathbf{v}t)\, d^3v = m\rho_0 \exp\left(-\frac{k^2 k_B T}{2m} t^2\right). \tag{17}$$

As an important insight from Eq. (17), the decay of positions-space density correlations in Eq. (16) is Gaussian not because of collisions or dissipation, but because gas particles with thermally distributed velocities stream out of phase. In plasma physics, such decay is broadly referred to as phase mixing [112 – 118].

The Fourier transform of the position-space correlation function in Eq. (16) gives the spectrum of density fluctuations,

$$S(k,\Omega) = \int_{-\infty}^{\infty} R_k(t)\exp(i\Omega t)dt = \frac{\rho_0 m}{k}\sqrt{\frac{2\pi m}{k_B T}}\exp\left(-\frac{m\Omega^2}{2k^2 k_B T}\right). \quad (18)$$

The collisionless Boltzmann equation thus leads to the Gaussian spectrum of spontaneous density fluctuations. This result is fully consistent with earlier numerical simulations [94 – 99, 107], where the Gaussian spectrum [Eq. (12)] emerges as the universal $y \to 0$ limit of the spectrum of spontaneous density fluctuations. Comparing Eq. (12) with the solution for the spectrum of spontaneous density fluctuations found in Eq. (18), we recognize the constant $\Omega_0$ in Eq. (12) as $\Omega_0^2 = 2k^2 k_B T/m = k^2 v_0^2$ .

**5. Stimulated scattering via free-streaming kinetics**

To understand the spectral properties of nonlinear light scattering from laser-driven density gratings, we examine the properties of perturbative solutions to the collisionless Boltzmann equation with an external force term as dictated by the laser-driven ponderomotive force in Eq. (3). With a laser field as defined by Eq. (1), the **K** projection of the spectral component of the force driving a density grating with frequency Ω and wave vector **K** is found from Eq. (3) as

$$\mathbf{F}_\Omega(\mathbf{r},\mathrm{t}) = F_\Omega(\mathbf{K}/K)i\exp[i(\mathbf{K}\cdot\mathbf{r} - \Omega t)] + \mathrm{c.c.} \quad (19)$$

with $F_\Omega = \alpha K E_p E_s^*$.

We now solve Eq. (13) for the distribution function $f(\mathbf{r},\mathbf{v},t)$ in the form of Eq. (10), with $f_1(\mathbf{r},\mathbf{v},t)$ understood as a small deviation from the Maxwellian equilibrium distribution,

$$f_0(\mathbf{r},\mathbf{v},t) = f_0(\mathbf{v}) = \frac{1}{\pi^{3/2} v_0^3}\exp\left(-\frac{v^2}{v_0^2}\right), \quad (20)$$

with thermal velocity $v_0 = \sqrt{2k_B T_0/m}$, equilibrium temperature $T_0$, and Boltzmann constant $k_B$.

Searching for $f_1(\mathbf{r},\mathbf{v},t)$ in the form of

$$f_1(\mathbf{r},\mathbf{v},t) = \delta f(\mathbf{K},\mathbf{v},\Omega)\exp[i(\mathbf{K}\cdot\mathbf{r} - \Omega t)], \quad (21)$$

transforms Eq. (13) to

$$-\Omega\delta\not{f} + Kv\delta\not{f} + \frac{\rho_0}{m}F_\Omega\frac{\partial\not{f}_0}{\partial v} = 0, \tag{22}$$

where $v = \mathbf{K}\cdot\mathrm{v}/K$, $\mathbf{v}_\perp = \mathbf{v} - v\,\mathbf{K}/K$, $\not{f}_0(v) = \int f_0(\mathbf{v})d^2v_\perp$, such that $\int \not{f}_0(v)\,dv = 1$, and $\delta\not{f} = \delta\not{f}(K, v, \Omega) = \int \delta f(\mathbf{K}, \mathbf{v}, \Omega)d^2v_\perp$

Solving Eq. (22) for $\delta\not{f}$, we find

$$\delta\not{f}(K, v, \Omega) = \frac{\rho_0 F_\Omega}{m(\Omega - Kv)}\frac{\partial\not{f}_0}{\partial v}. \tag{23}$$

The field-induced density change is now found as

$$\delta\rho_\Omega = \int \delta\not{f}(K, v, \Omega)dv = \frac{\rho_0}{m}\int \frac{F_\Omega}{\Omega - Kv}\frac{\partial\not{f}_0}{\partial v}dv. \tag{24}$$

To relate these equations to the nonlinear polarization and nonlinear susceptibility [Eqs. (5), (6)], we use Eq. (24) to express the complex conjugate of $\delta\rho_\Omega$ as

$$\delta\rho_\Omega^* = \alpha\,[\mathcal{q}(K, \Omega)]^* E_p^* E_s, \tag{25}$$

with

$$\mathcal{q}(K, \Omega) = \frac{\rho_0}{m}K\int_{-\infty}^{\infty}\frac{\partial\not{f}_0/\partial v}{\Omega - Kv}dv \tag{26}$$

The nonlinear polarization behind stimulated scattering [Eq. (6)] can now be written as

$$P^{(3)}(\omega_s) = \frac{\alpha^2\mathcal{N}}{\rho_0}[\mathcal{q}(K, \Omega)]^*\left|E_\mathrm{p}\right|^2 E_\mathrm{s}. \tag{27}$$

The pertinent nonlinear susceptibility is

$$\chi^{(3)}(\omega_s) = \frac{\alpha^2\mathcal{N}}{\rho_0}[\mathcal{q}(K, \Omega)]^*. \tag{28}$$

The integral in Eq. (26) is taken via the Sokhotski–Plemelj theorem [120, 121],

$$\lim_{\epsilon\to 0}\frac{1}{x + i\epsilon} = \mathcal{P}\frac{1}{x} - i\pi\delta(x), \tag{29}$$

with $\mathcal{P}$ standing for the Cauchy principal-value integral, leading to

$$\mathrm{Im}\left[\chi^{(3)}\right] = -\frac{\alpha^2\mathcal{N}}{k_B T}\sqrt{\pi}\theta\exp(-\theta^2), \tag{30}$$

with $\theta = \frac{\Omega}{Kv_0}$.

The stimulated scattering gain is now found from Eq. (8), yielding

$$g_s(\theta) = \frac{8\pi^{5/2}\omega_s}{n^2c^2}\frac{\alpha^2\mathcal{N}}{k_B T}\theta\exp(-\theta^2), \tag{31}$$

or, in terms of the electrostrictive coefficient $\gamma_e$,

$$g_s(\theta) = \frac{\sqrt{\pi}\omega_s}{n^2c^2}\frac{{\gamma_e}^2}{\rho_0 v_0^2}\theta\exp(-\theta^2). \tag{32}$$

Eqs. (23) – (26) reveal the significance of the $(\Omega - Kv)^{-1}$ pole for the properties of nonlinear scattering from laser-driven density gratings in a collisionless gas. Integrals with poles of this type have been studied in the vast literature on plasma physics focusing on plasma dispersion, Landau damping, and van Kampen plasma modes in collisionless plasmas [108 – 119]. The physical setting examined in our study here is, clearly, distinctly different, as the electric neutrality of the gas in our treatment makes plasma dispersion and collective plasma oscillations irrelevant. Instead, the $(\Omega - Kv)^{-1}$ pole in our study is a signature of the $\Omega = Kv$ kinetic resonance, highlighting the special role of gas particles with velocity $\mathbf{v}$ meeting $\mathbf{K} \cdot \mathbf{v} = \Omega$. Such particles stay in phase with the driver force $\mathbf{F}_\Omega(\mathbf{r}, \mathrm{t})$ [Eq. (19)], providing a resonant contribution to the laser-induced change in gas density, as expressed by the $(\Omega - Kv)^{-1}$ denominator in Eq. (24).

Figure 1a illustrates the ballistic characteristics $Kvt - \Omega t$ of free-streaming gas particles moving with a velocity $v < \Omega/K$ (green line), $v = \Omega/K$ (blue line), and $v > \Omega/K$ (red line) versus the phase $Kx - \Omega t$ of the gas-density grating. As is readily seen from this figure, gas particles that see the phase $\Phi_0$ of the density grating at the moment of time $t = 0$ (the point where all the $Kvt - \Omega t$ lines start in Fig. 1a) overtake the $\Phi_0$ phase front if their velocity $v$ is larger than $\Omega/K$ (red line in Fig. 1a), lag behind this front if $v < \Omega/K$ (cyan line), or stay in phase with the grating if $v = \Omega/K$ (blue line).

The effect that such phase mixing has on the time evolution of a gas-density grating is illustrated in Fig. 1b, which shows how the perturbation $f_1(x, v, t)$ of the distribution function $f(x, v, t)$ evolves in time in the frame of reference moving with the density grating. Given that, at time $t = 0$, the modulation of this perturbation is $f_1(x, v, t = 0) = \delta\rho(x = 0, t = 0) f_0(v) \cos(Kx)$ for particles with velocity distribution $f_0(v)$ (the upper left panel in Fig. 1b), the evolution of $f_1(x, v, t)$ with time is found as $f_1(x, v, t) = \delta\rho(x = 0, t = 0) f_0(v) \cos[K(x - vt)]$. The free streaming of gas particles thus shears the fringes of $f_1(x, v, t)$ in the $(x, v)$ phase space. The middle and right panels in the upper part of Fig. 1b illustrate this effect by showing $f_1(x, v, t)$ at two moments of time, $t = \tau_0$ and $t = 3\tau_0$, with $\tau_0 = (Kv_0)^{-1}$ and with color encoding $v_0 f_1(x, v, t)/\delta\rho(0,0)$.

Perturbation of the distribution function $f_1(x, v, t)$ translates into a change in the gas density, $\delta\rho(x, t) = \int_{-\infty}^{\infty} f_1\,(x, v, t)\, dv = \delta\rho(0,0) \int_{-\infty}^{\infty} f_0\,(v) \cos(Kx - Kvt)\, dv$. Thus, for each $x$ and $t$, $\delta\rho(x, t)$ is given by an integral of $f_1(x, v, t) = \delta\rho(x = 0, t = 0) f_0(v) \cos[K(x - vt)]$, shown by color in the upper panels in Fig. 1b, over all the velocities, i.e., along the vertical lines in the upper panels in Fig. 1b. Over time, $f_1(x, v, t)$, appearing as the integrand in the solution for $\delta\rho(x, t)$, becomes a rapidly oscillating function of $v$ (e.g., at $t = 3\tau_0$ in Fig. 1b). As the integration in $v$ in the expression for $\delta\rho(x, t)$ extends over these oscillations, the modulation amplitude of $\delta\rho(x,t)$ decreases. This tendency is illustrated by the lower panels in Fig. 1b. For Maxwellian velocity distribution $f_0(v) = \left(\sqrt{\pi}v_0\right)^{-1}\exp(-v^2/v_0^2)$, $\delta\rho(x,t)$ is found analytically, $\delta\rho(x,t) = \delta\rho(0,0)\cos(Kx)\exp[-(Kv_0t)^2/4]$. The contrast of density-gradient modulation in this setting falls off as $\exp[-t^2/(2\tau_0)^2]$ with a time constant $2\tau_0 = 2(Kv_0)^{-1}$ (the lower panels in Fig. 2b).

**6. The spectrum of coherent scattering: causality and Kramers–Kronig relations**

While the stimulated gain spectrum $g_s(\Omega)$ is defined by the imaginary part of the cubic susceptibility $\chi^{(3)}(\Omega)$ [Eq. (8)], the spectral lineshape $Q(\Omega)$ of the coherent signal generated via scattering from the laser-driven density grating is determined by the spectrum of $\left|\chi^{(3)}(\Omega)\right|^2 = \left\{\mathrm{Re}\left[\chi^{(3)}(\Omega)\right]\right\}^2 + \left\{\mathrm{Im}\left[\chi^{(3)}(\Omega)\right]\right\}^2$. To find $Q(\Omega)$, we observe that, as one of the inherent, readily verifiable [119 – 121] properties of the Sokhotski–Plemelj expansion [Eq. (29)], the real and imaginary parts of the susceptibility $\chi^{(3)}$ defined by Eqs. (26) – (29) are manifestly causal as they satisfy the Kramers–Kronig relations,

$$\mathrm{Re}\left[\chi^{(3)}(\omega)\right] = -\mathcal{H}\left\{\mathrm{Im}\left[\chi^{(3)}\right]\right\} = \pi^{-1}\mathcal{P}\int_{-\infty}^{\infty}(\omega'-\omega)^{-1}\mathrm{Im}\left[\chi^{(3)}(\omega')\right]d\omega', \quad (33)$$

$$\mathrm{Im}\left[\chi^{(3)}(\omega)\right] = \mathcal{H}\left\{\mathrm{Re}\left[\chi^{(3)}\right]\right\} = -\pi^{-1}\mathcal{P}\int_{-\infty}^{\infty}(\omega'-\omega)^{-1}\mathrm{Re}\left[\chi^{(3)}(\omega')\right]d\omega', \quad (34)$$

where

$$\mathcal{H}\{f\} = \pi^{-1}\mathcal{P}\int_{-\infty}^{\infty}(\omega-\omega')^{-1}f(\omega')d\omega' \quad (35)$$

is the Hilbert transform

With $\mathrm{Im}\left[\chi^{(3)}(\Omega)\right]$ as found in Eq. (30), $\mathrm{Re}\left[\chi^{(3)}(\Omega)\right]$ can thus be computed via Eq. (33).

For deeper insights into the properties of $Q(\Omega)$, we express the solution for $\mathrm{Im}\left[\chi^{(3)}(\Omega)\right]$ found in Eq. (30) in terms of the Faddeeva function [122],

$$w(\xi) = \exp(-\xi^2)\,\mathrm{erfc}(-i\xi) = \exp(-\xi^2)\left[1 + \left(2i/\sqrt{\pi}\right)\int_0^{\xi}\exp(\eta^2)\,d\eta\right], \quad (36)$$

as

$$\mathrm{Im}\left[\chi^{(3)}\right] = \frac{\alpha^2\mathcal{N}}{2k_BT}\sqrt{\pi}\,\mathrm{Re}[w'(\theta)]. \quad (37)$$

The stimulated gain [Eqs. (31), (32)] can now be expressed as

$$g_s(\theta) = -\frac{4\pi^{5/2}\omega_s}{n^2c^2}\frac{\alpha^2\mathcal{N}}{k_BT}\mathrm{Re}[w'(\theta)]. \quad (38)$$

The inverse Hilbert transform of $\mathrm{Im}\left[\chi^{(3)}\right]$ in Eq. (33) now gives

$\mathrm{Re}\left[\chi^{(3)}(\Omega)\right] = \frac{\alpha^2 \mathcal{N}}{2k_B T}\sqrt{\pi}\mathrm{Im}[w'(\theta)]\,.$ (39)

Combining Eqs. (37) and (39), we find for the full complex cubic susceptibility: $\chi^{(3)}(\Omega) = \frac{\alpha^2 \mathcal{N}}{2k_B T}\left[1 - i\sqrt{\pi}\theta \mathrm{w}(-\theta)\right] = -\frac{i\sqrt{\pi}\alpha^2 \mathrm{N}}{2k_B T}\mathrm{w}'(-\theta)$. The lineshape of the coherent signal is thus given by

$Q(\theta) = \left\{\mathrm{Re}\left[\chi^{(3)}(\theta)\right]\right\}^2 + \left\{\mathrm{Im}\left[\chi^{(3)}(\theta)\right]\right\}^2 = \frac{\pi\alpha^4 \mathcal{N}^2}{4(k_B T)^2}|w'(\theta)|^2.$ (40)

Eqs. (38) and (40) define the map whereby the Gaussian lineshapes inherent in dilute-gas spontaneous scattering [Eq. (18)] transform into the lineshapes of stimulated and coherent scattering (Table 1). These equations also provide physically intuitive, closed-form solutions for the spectra of stimulated and coherent light scattering from laser-driven gratings in dilute gases, expressing them in terms of the first derivative of the Faddeeva function $w'(\Omega)$. Specifically, Eq. (38) expresses the spectrum $g_s(\Omega)$ of stimulated gain in terms of the real part of $w'(\Omega)$. The spectral profile $Q(\Omega)$ of the coherent signal, on the other hand, is defined by $|w'(\Omega)|^2$ [Eq. (40)].

In Fig. 2, the spectral shapes $g_s(\theta)$ and $Q(\theta)$ of stimulated and coherent scattering are presented alongside the spectrum $S(\theta)$ of spontaneous dilute-gas density fluctuations [Eq. (18)]. As is readily seen from Eqs. (18), (38), and (40), the bandwidth covered by all three spectral profiles is on the order of $\Omega_0 = kv_0$ (Table 1), thus connecting to the time $\tau_0$ required for free-streaming particles with a thermal velocity $v_0$ to traverse a distance of $k^{-1}$, $\Omega_0 = \tau_0^{-1}$. This is in a stark contrast with the spectra of spontaneous and stimulated Brillouin scattering, whose bandwidths connect to the inverse phonon lifetime [10 – 15] and, hence, to the shear and dilation viscosity properties of the medium [12].

As is readily seen from Fig. 2, the $S(\theta)$ and $Q(\theta)$ lineshapes are very close to each other for small and moderate $\theta$. Indeed, their small-$\theta$ expansions, $S(\theta) \approx 1 - \theta^2 + \theta^4/2$ and $Q(\theta) \approx 1 - (4 - \pi)\theta^2 + (20/3 - 2\pi)\,\theta^4/2$, deviate by only $(\pi - 3)\theta^2 \approx 0.14\theta^2$ as the leading correction term. The full width at half maximum (FWHM) of $Q(\theta)$ is only ≈ 9% larger than the FWHM of $S(\theta)$. For large $\theta$, however, the behavior of $S(\theta)$ and $Q(\theta)$, as is also readily seen in Fig. 2, is drastically different. The large-$\theta$ decay of $Q(\theta)$ is significantly slower, $w'(\theta) \sim -i/\left(\sqrt{\pi}\theta^2\right)$, $Q(\theta) \sim 1/(4\theta^4)$, as compared to the large-$\theta$ decay of $S(\theta)$, $S(\theta) \sim \exp(-\theta^2)$. The properties of extreme-value statistics of $S(\theta)$ and $Q(\theta)$ are thus distinctly different, as they converge to different extreme-value distributions [123, 124]. Yet, the decay $Q(\theta) \sim 1/(4\theta^4)$ of the spectrum of the coherent signal is fast enough for the root-mean-square linewidth of $Q(\theta)$ to remain finite, $(\Delta\theta)_Q = \sqrt{3}/2$, and even comparable, $(\Delta\theta)_Q/(\Delta\theta)_S = \sqrt{3/2} \approx 1.22$, to the root-mean-square linewidth of $S(\theta)$, $(\Delta\theta)_S = 1/\sqrt{2}$.

**7. Fluctuation – dissipation relation**

While the $S(\Omega)$ and $Q(\Omega)$ lineshapes are both symmetric with respect to $\Omega = 0$ (Fig. 1, Table 1), $g_s(\Omega)$ is an odd function of $\Omega$, antisymmetric relative to $\Omega = 0$ (Fig. 1, Table 1), as it is defined by the negative derivative of $S(\Omega)$. The gain $g_s$ at the Stokes frequency thus becomes the loss of equal magnitude at the anti-Stokes frequency, $g_s(-\Omega) = -g_s(\Omega)$. This property of $g_s(\Omega)$ is fully consistent with causality, as expressed by the Kramers – Kronig relations [Eqs. (33), (34)], as well as with the available measurements [34 – 36, 125 – 127]. Moreover, this behavior of $g_s(\Omega)$ can be viewed as a manifestation of the fundamental connection [128 – 138] between fluctuations and dissipation in a gas close to its equilibrium. To gain insights into the inner workings enabling this connection in nonlinear light scattering from laser-driven density gratings, we resort to the solution to the collisionless Boltzmann equation for the field-induced density change [Eqs. (24) – (26)] and recognize $\mathcal{q}(K,\Omega)$ in Eq. (26) as the spectrum of the linear response to the ponderomotive force [Eqs. (3), (19)]. Taking the integral in Eq. (26) via the Sokhotski–Plemelj formula [Eq. (29)], we find

$$\mathrm{Im}[\mathcal{q}(K,\Omega)] = -\pi\mathcal{N}(\partial f_0/\partial v)_{v=\Omega/K}. \tag{41}$$

For Maxwellian $f_0(v) = \pi^{-1/2}v_0^{-1}\exp(-v^2/v_0^2)$, Eq. (41) gives

$$\mathrm{Im}[\mathcal{q}(K,\Omega)] = \frac{2\rho_0\Omega}{Kmv_0^3}\sqrt{\pi}\exp\left(-\frac{\Omega^2}{K^2v_0^2}\right). \tag{42}$$

Comparing this result to Eq. (18), with a variable $k$ of spontaneous density fluctuations in Eqs. (16) – (18) identified with the wave number $K$ of the laser-induced density grating in stimulated scattering, we find

$$m\mathrm{Im}[\mathcal{q}(K,\Omega)] = \frac{\Omega}{2k_BT}S(K,\Omega). \tag{43}$$

Eq. (43) is an expression of the fluctuation – dissipation theorem. In the context of nonlinear light scattering, this equation establishes that the change in gas density in response to the laser-driven ponderomotive force [Eqs. (3), (19)] is intrinsically related to spontaneous density fluctuations. More specifically, the relaxation of density fluctuations in a gas perturbed out of its equilibrium with a laser driver must follow the same decay law as the decay of spontaneous density fluctuations at equilibrium. The Stokes gain (or anti-Stokes loss) in stimulated light scattering by laser-induced density gratings is the dissipation counterpart to spontaneous density fluctuations. The stimulated scattering gain spectrum $g(\Omega)$ thus directly connects to and can be found from the spectrum of spontaneous density fluctuations $S(\Omega)$ (Table 1). The distinctive

$$g_s(\Omega) \propto \Omega/(2k_BT)\exp\left[-\Omega^2/(Kv_0)^2\right] \quad (44)$$

motif in the stimulated gain spectrum [Eq. (31)], giving rise to the asymmetry of $g_s(\Omega)$ relative to $\Omega = 0$, is thus a readily recognizable signature of fluctuation – dissipation connection, $g_s(\Omega) \propto \Omega/(2k_BT)\,S(K,\Omega)$.

## 8. Benchmarking against experimental data

As a demonstration of practical value of the closed-form solutions derived in this study for the spectral shapes of dilute-gas nonlinear scattering, we benchmark Eqs. (31), (32), (38) against the gold-standard data – the stimulated gain spectra measured by She et al. [35] and Mironov et al. [93]. The results of such benchmarking are presented in Figs. 3a and 3b.

Figure 3a compares the spectral profile of stimulated gain $g_s(\Omega)$ calculated via Eqs. (31), (32), (38) with the stimulated gain spectrum measured by She et al. [35] for argon at a pressure of 0.9 atm and a temperature of 296 K. The pump and Stokes beams in these experiments were delivered by a fixed-frequency dye laser operating at a 585 nm and a tunable dye laser, crossing at an angle of about 179°. For argon, $m \approx 6.63\cdot10^{-23}$ g, at such a pressure and temperature, $v_0 \approx 350$ m/s, leading to $\nu_c \approx 4.06\cdot10^9$ s$^{-1}$. For the laser wavelength and beam-crossing geometry of this experiment, $K \approx 2.15\cdot10^7$ m$^{-1}$, $Kv_0 \approx 7.54\cdot10^9$ s$^{-1}$, and $\Omega_0/(2\pi) \approx 1.2$ GHz. Stimulated light scattering under these conditions is, to a satisfactory approximation, collisionless, with $y \approx 0.54$. The expected peaks of the stimulated gain and loss spectra are at $\pm\Omega_s/\left(2\sqrt{2}\pi\right) \approx \pm 0.84$ GHz.

Presented in Fig. 3b is the spectral profile of stimulated gain $g_s(\Omega)$ as predicted by Eqs. (31), (32), (38) vis-à-vis the stimulated gain spectrum measured by Mironov et al. [93] for krypton at a pressure of 1.0 atm and a temperature of about 300 K. The pump and Stokes beams in these studies were produced by frequency-quadrupling the 1064-nm output of a Nd: YAG laser, yielding radiation with a wavelength of about 266 nm, and were crossed at an angle of about 175°. For krypton, $m \approx 1.39\cdot10^{-22}$ g, at such a pressure and temperature, $v_0 \approx 240$ m/s, leading to $\nu_c \approx 4.0\cdot10^9$ s$^{-1}$. For the laser wavelength and beam-crossing geometry of this experiment, $K \approx 4.7\cdot10^7$ m$^{-1}$, translating into a grating period $l_g = 2\pi/K \approx 133$ nm, $Kv_0 \approx 1.15\cdot10^{10}$ s$^{-1}$, and $\Omega_0/(2\pi) \approx 1.83$ GHz. Stimulated light scattering under these conditions is largely collisionless, with $y \approx 0.34$. The expected peak of the stimulated gain spectrum is at $\Omega_s/(2\sqrt{2}\pi) \approx 1.3$ GHz.

Eqs. (31), (32), (38) are seen to provide a highly accurate description of dilute-gas stimulated gain spectra in both experiments. Successful fitting of the experimental data in Figs. 3a and 3b has already been demonstrated in earlier studies [35, 93], but was based on different premises. Stimulated gain and loss spectra for krypton at 1.0 atm were accurately fitted [35] by combining the Yip – Nelkin framework for the calculation of the spectrum $S(k,\Omega)$ of spontaneous density fluctuations with the fluctuation – dissipation relation, which, as shown in Section 7, is expressible in the form of Eqs. (43) and (44). The Yip – Nelkin framework finds $S(k,\Omega)$ by solving the linearized Boltzmann equation with zero external force term [Eqs. (9), (11) with $\mathbf{F} = 0$] and with the collision term $C[f]$ taken in the linear BGK approximation [94 – 106]. This framework is distinctly different from the approach that we take in this study by solving the linearized Boltzmann equation (13) with zero collision term $C[f]$, but with a nonzero external force term as dictated by the electrostrictive force $\mathbf{F}(\mathbf{r}, \mathrm{t})$ in Eq. (3). Both approaches should converge, however, in the parameter space where the fluctuation – dissipation relation in the form of Eqs. (43), (44) is applicable and where collisions are negligible. Both requirements are achieved in dilute-gas experiments on stimulated light scattering in krypton at 1 atm.

Stimulated gain spectra of argon at moderate pressures, on the other hand, were successfully fitted [93] by solving the linearized Boltzmann equation (13) with an electrostrictive force term $\mathbf{F}(\mathbf{r}, \mathrm{t})$ as defined by Eq. (3) and with the collision term taken in the linear BGK approximation. This approach should also converge to the approach adopted in this study in the dilute-gas limit, i.e., in the regime of low $y$.

Thus, whenever the fluctuation – dissipation relation in the form of Eqs. (43), (44) is applicable and collisions are negligible, both approaches that have earlier enabled an accurate fitting of stimulated gain spectra [35, 93] converge to the approach that we developed in this study. Both requirements are achieved, even though not with a large margin, in dilute-gas experiments on stimulated light scattering in argon and krypton considered here. We therefore do not expect that the approach developed in this paper will provide a higher accuracy of fitting of experimental stimulated gain spectra. The value of the developed approach, instead, is that it reveals the universality of spectral shapes in dilute-gas stimulated and coherent scattering. Indeed, when plotted as functions of the dimensionless variable $\theta = \Omega/(Kv_0)$, the stimulated gain/loss spectra of argon and krypton in Figs. 3a and 3b, as well as any dilute-gas stimulated gain/loss spectrum for that matter, acquire the form of the universal spectral profile $g_s(\theta)$ in Fig. 2. The developed approach leads to a physically intuitive solution for such spectra, expressing them in a manifestly causal closed form that satisfies the Kramers – Kronig relations and fulfills the fluctuation – dissipation theorem. As one of its merits, the developed framework helps avoid misattributing the bandwidth of dilute-gas nonlinear scattering spectra to phonon lifetimes as it reveals the true physical content of this bandwidth as an expression and a quantifier of the phase-mixing time, i.e., the time required for free-streaming gas particles with a mean thermal velocity to traverse the distance equal to the spacing of the laser-driven density grating.

## 9. Conclusion

To summarize, we have identified universal spectral profiles inherent in stimulated and coherent light scattering from laser-driven dilute-gas density gratings and derived physically intuitive closed-form solutions for such spectral shapes, expressing them in terms of the Faddeeva function. When cast in this form, the spectra of nonlinear scattering are manifestly causal, satisfying the Kramers – Kronig relations and fulfilling the fluctuation – dissipation theorem. Benchmarking against gold-standard experiments on dilute-gas stimulated light scattering verifies a high descriptive and explanatory power of these solutions and demonstrates their ability to isolate universality motifs in nonlinear scattering spectra as the key to fully understanding the physics behind dilute-gas nonlinear light scattering. As one of important outcomes, such analysis reveals the physical content of the bandwidth of dilute-gas nonlinear scattering spectra as an expression and a quantifier of the phase-mixing time, i.e., the time required for free-streaming gas particles with a mean thermal velocity to traverse the distance equal to the spacing of the laser-driven density grating.

## Acknowledgments

This research was supported in part by the DOE Office of Science through the DOE Inertial Fusion Energy Science and Technology Accelerator Research (IFE-STAR) program (grant # DE-SC0024882).

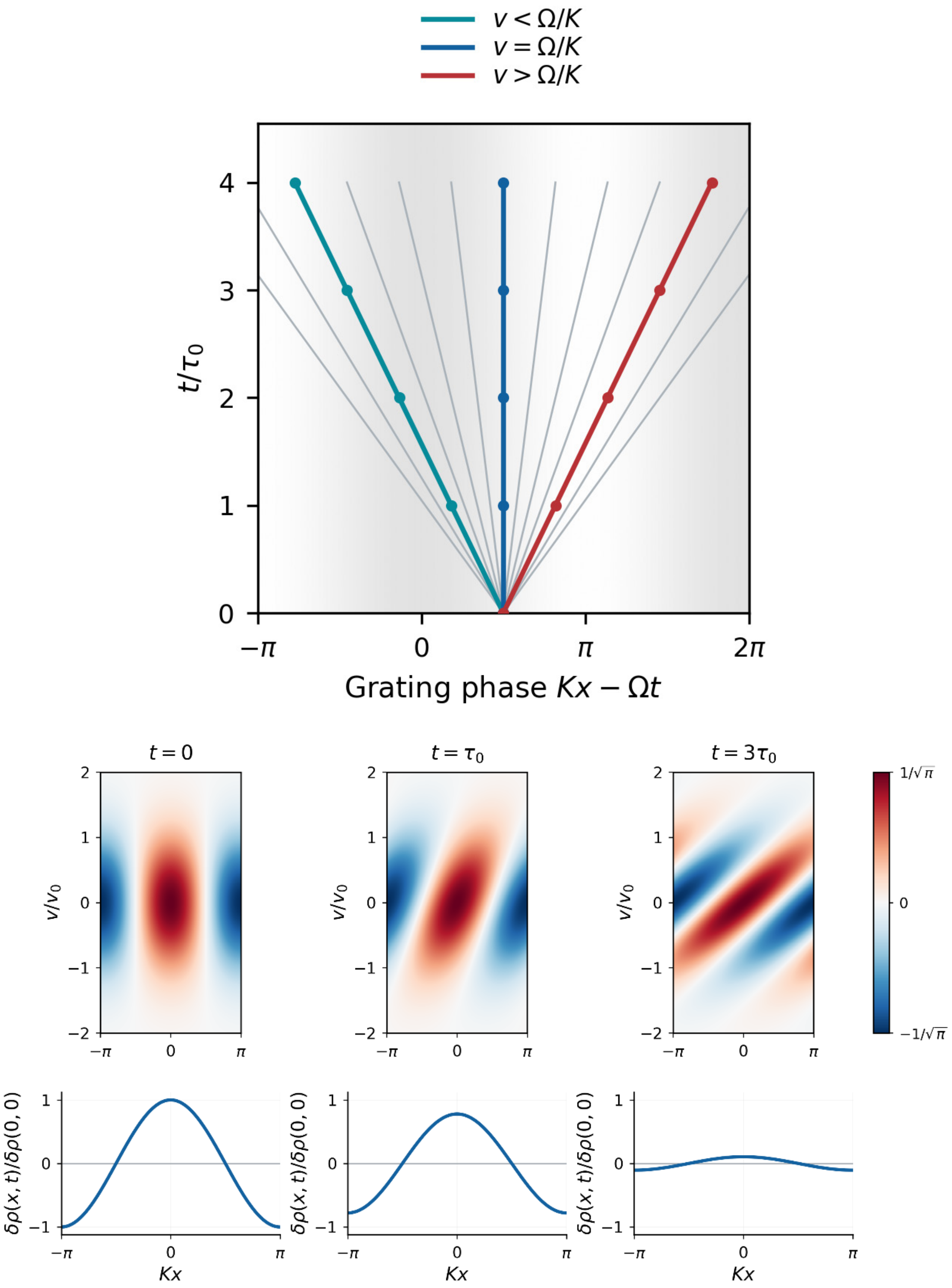

Fig. 1. (a) Ballistic characteristics $Kvt - \Omega t$ of free-streaming gas particles moving with a velocity $v < \Omega/K$ (green line), $v = \Omega/K$ (blue line), and $v > \Omega/K$ (red line) versus the phase $Kx - \Omega t$ of the gas-density grating. (b) Perturbation $f_1(x, v, t)$ of the gas-particle distribution function (upper panels) and the respective change in the gas density $\delta\rho(x, t)$ (lower panels) at three moments of time, $t = 0$, $t = \tau_0$, and $t = 3\tau_0$, with $\tau_0 = (Kv_0)^{-1}$. Color encodes $v_0 f_1(x, v, t)/\delta\rho(0,0)$.

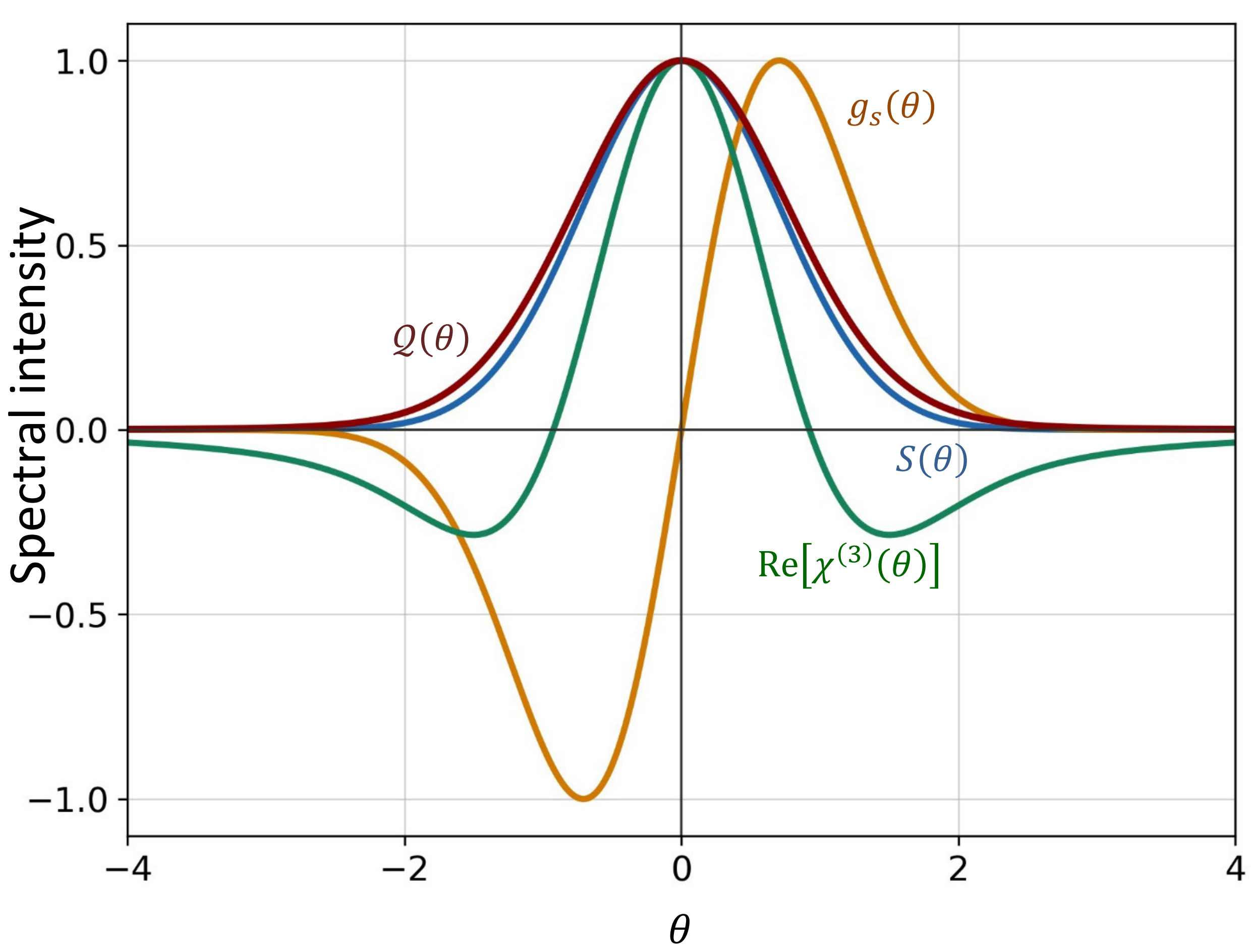

Fig. 2. Spectral lineshapes in light scattering from laser-driven density gratings: (blue) the spectrum of spontaneous density fluctuations, (brown) the spectrum of stimulated gain and loss, and (red) the spectrum of coherent scattering. Shown in green is the spectrum of the real part of the cubic susceptibility. All the lineshapes are normalized to their maximal values and are plotted as functions of $\theta = \Omega/(Kv_0)$.

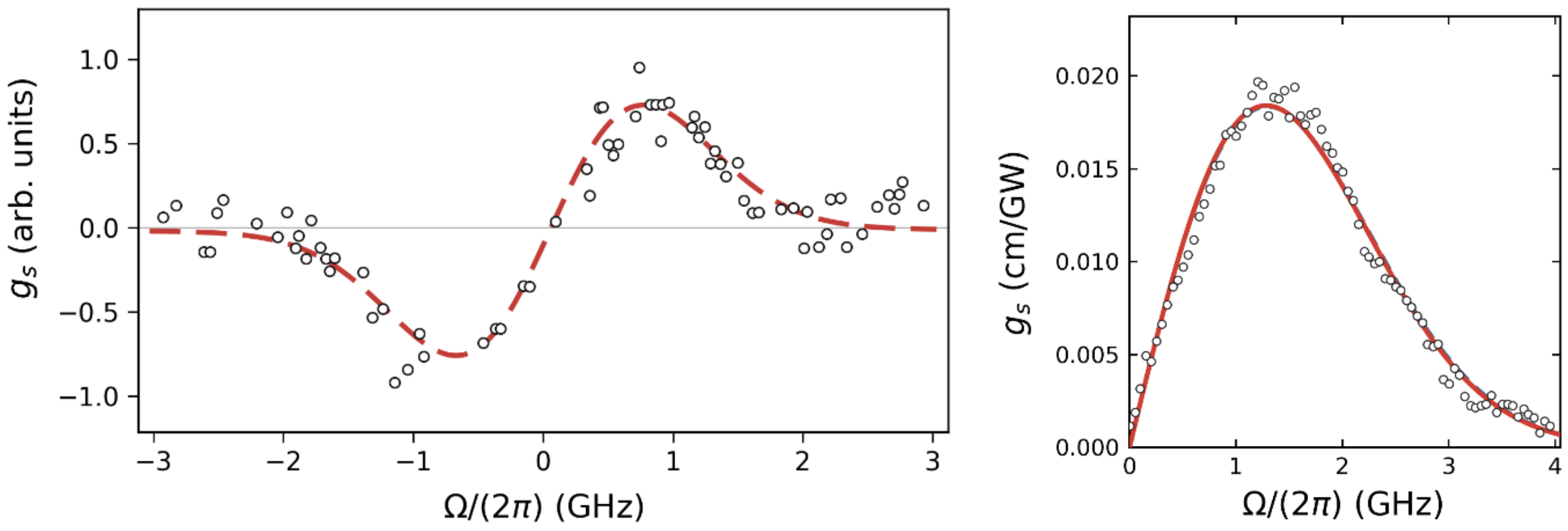

Fig. 3. Stimulated gain spectra $g_s(\Omega)$ calculated by using Eqs. (31), (32), (38) benchmarked against (a) the stimulated gain/loss spectrum measured by She et al. [35] for argon at a pressure of 0.9 atm and (b) the stimulated gain spectrum measured by Mironov et al. [93] for krypton at 1.0 atm: (red lines) calculations using Eqs. (31), (32), (38), (circles) experimental data from She et al. [35] (a) and Mironov et al. [93] (b).

Table 1. Universal collisionless spectral lineshapes in light scattering from laser-driven density gratings

| Process | Spontaneous scattering | Stimulated scattering | Coherent scattering |
|---|---|---|---|
| Lineshape | Gaussian, Eq. (18) | The real part of the derivative of the Faddeeva function, Eq. (38) | Modulus squared of the derivative of the Faddeeva function, Eq. (40) |
| Central motif | $S(k,\Omega) \propto \exp\left(-\frac{m\Omega^2}{2k^2 k_B T}\right)$ | $g_s(\Omega) \propto -\mathrm{Re}[w'(\Omega)]$ | $Q(\Omega) \propto \lvert w'(\Omega)\rvert^2$ |
| Parity | $S(k,\Omega) = S(k,-\Omega)$ | $g_s(-\Omega) = -g_s(\Omega)$ | $Q(\Omega) = Q(-\Omega)$ |
| Linewidth | inverse phase-mixing time, $\tau_0^{-1} = K v_0$ | | |
| Kramers–Kronig relations | | $\mathrm{Re}\left[\chi^{(3)}(\omega)\right] = \pi^{-1}\mathcal{P}\int_{-\infty}^{\infty}(\omega'-\omega)^{-1}\mathrm{Im}\left[\chi^{(3)}(\omega')\right]d\omega'$ <br> $\mathrm{Im}\left[\chi^{(3)}(\omega)\right] = -\pi^{-1}\mathcal{P}\int_{-\infty}^{\infty}(\omega'-\omega)^{-1}\mathrm{Re}\left[\chi^{(3)}(\omega')\right]d\omega'$ | |
| Fluctuation – dissipation connection | $g_s(\Omega) \propto \Omega/(2k_B T)\, S(K,\Omega)$ | | |